\documentclass[a4paper]{jpconf}
\usepackage{makecell}
\usepackage{graphicx}
\usepackage{pifont}
\usepackage{float}
\usepackage{amsmath}
\begin{document}
\title{Interference filter based external-cavity diode laser with combined dual interference filters and largely adjustable  feedback range}

\author{Rui Chang$^{1}$, Jun He$^{1,2}$, and Junmin Wang$^{1,2,*}$}

\address{$^{1}$ State Key Laboratory of Quantum Optics and Quantum Optics Devices, and Institute of Opto-Electronics, Shanxi University, Taiyuan 030006, China\\
$^{2}$ Collaborative Innovation Center of Extreme Optics, Shanxi University, Taiyuan 030006, China\\
$^{*}$ Author to whom any correspondence should be addressed}

\ead{wwjjmm@sxu.edu.cn}
\hspace*{\fill}

\newcommand\keywords[1]{\textbf{Keywords}: #1}
\qquad\quad\quad\quad\keywords{external-cavity diode laser, combination of narrow-band interference\\\qquad\quad\quad\qquad filters, cateye configuration, tunable feedback portion, laser linewidth}

\begin{abstract}
External-cavity diode lasers (ECDL) are widely used as light sources in laser spectroscopy, atomic physics, and quantum optics. This study demonstrated a home-made 852-nm ECDL with variable feedback, using the combined dual narrow-band interference filters (IFs) as the laser longitudinal mode selection element. The combination of narrow-band IFs, mainly for the current commercially available narrow-band IFs with the full width at half maximum (FWHM) of approximately 0.5 nm for 780-895 nm. We designed different narrow-band IFs combinations to achieve a narrower FWHM, and applied them in the ECDL. The combination of the dual narrow-band IFs further reduces the laser linewidth. We measured the laser linewidth using a high-finesse Fabry-Perot cavity with a linewidth of approximately 10 kHz. The laser linewidth is approximately 176 kHz for the Single-IF-ECDL and 96 kHz for the Dual-IF-ECDL with the same feedback. In addition, we have experimentally verified the results of narrower laser linewidth and larger tuning range with increase in the feedback. The developed Dual-IF-ECDL has the capability of narrow linewidth and wavelength tunability and can be applied to precision spectroscopy, cooling and trapping of neutral atoms.
\end{abstract}

\section{Introduction}
Diode lasers offer the advantages of small size, low energy consumption, high efficiency, long life, and fast modulation, Therefore, they are widely used in laser communication, optical storage, laser medicine, lidar, laser ranging, and other fields \cite{wieman1991using, meiners2009multi, schiller2009einstein}. With the development of diode lasers and the demand for related fields, narrow linewidth, and wavelength tunability have gradually become the development direction of diode lasers. They are usually achieved by integrating the laser diode into an external cavity to form an external cavity diode laser (ECDL)\cite{cui2022advances}. 

In 1980, Lang and Kobayashi \cite{lang1980external} applied the external cavity feedback technique to diode lasers, and the added external cavity can make the laser linewidth narrowed, and the tuning of the wavelength can be achieved by adjusting the length of the external cavity. In 1981, Fleming and Mooradian \cite{fleming1981spectral} first used a diffraction grating to feedback a part of the output laser of the diode laser into the active region, thereby reducing the linewidth of the laser to 1.5 MHz. Two types of diode lasers that typically use a diffraction grating as an external cavity are the Littrow \cite{arnold1998simple,hawthorn2001littrow}  and Littman \cite{harvey1991external} configurations. These lasers are sensitive to acoustic and mechanical interferences, particularly when spring-loaded kinematic mount is used to align the grating or feedback optics that require precise alignment. Therefore, more structurally stable external-cavity diode lasers are required. In 1988 Zorabedian and Trutna \cite{zorabedian1988interference} experimentally used an external-cavity laser with an interference filter (IF) as the wavelength-selective element, which has approximately the same wavelength tuning range as the conventional grating external-cavity laser. Baillard $\emph{et al}$ \cite{baillard2006interference} showed both theoretically and experimentally that the interference filter external-cavity diode lasers (IF-ECDL) are more stable. Gilowski $\emph{et al}$ \cite{gilowski2007narrow} experimentally investigated a new laser design based on a self-seeded taper amplifier with wavelength stabilization using narrow-band high-transmission IFs to manipulate neutral atoms. Ruan Jun $\emph{et al}$ \cite{ruan2010robust} developed a highly stable IF-ECDL with a wavelength of 852 nm for atomic clock experiments. Martin $\emph{et al}$ \cite{martin2016external} developed an IF-ECDL using dual IFs as longitudinal-mode selection elements. Vassiliev $\emph{et al}$ \cite{vassiliev2019vibration}  presented a highly stable IF-ECDL and offered the possibility of using an ECDL for airborne systems. 2020 Linbo Zhang $\emph{et al}$ \cite{2020Development} developed an IF-stabilized external-cavity diode laser for space applications. 2022 Lingqiang Meng $\emph{et al}$ \cite{meng2022design} developed IF-ECDL to pass the aerospace environmental tests, indicating that the IF-ECDL is suitable for space missions in the China Space Station (CSS).

A diode laser with an IF differs from a diode laser with a grating external cavity \cite{Liu2018852nm}. The grating of the diode laser with a grating is both an external feedback element and a mode-selecting element and is susceptible to external mechanical perturbations. The external cavity of the diode laser with an IF consists of the rear surface of the laser diode and cat-eye reflector. Different incidence angles of the IF correspond to different central wavelengths, and the wavelength selection is achieved by rotating the IF, the feedback is achieved by the cat-eye reflector, and the cat-eye reflector is self-aligning and insensitive to intra-cavity optical misalignment. The robust alignment stability ensures an inherently good mode match between the external cavity and laser diode, so the overall stability of the diode laser with an IF is higher \cite{thompson2012narrow}. Moreover, in the case of ECDL with the interference filter, because the plane mirror provides the optical feedback, the structure can be designed in the ECDL to change the optical feedback without being limited by the intrinsic diffraction efficiency of the grating.

This study developed an ECDL with variable feedback, using the combined dual narrow-band IFs as the laser longitudinal mode selection element, and rotating the IF to tune the wavelength without changing the direction of the output laser. We achieved narrower full-width at half maximum (FWHM) results by designing two different narrow-band IFs combinations and applying them to the ECDL. In the experiment, the linewidth of the IF-ECDL output laser was compared for different cases, and how to combine the two IFs to achieve the optimal value of the laser linewidth and power. Finally, we measured the laser linewidth and wavelength tuning range of the IF-ECDL with different feedbacks, as the feedback increased, the laser linewidth decreased and the tuning range increased \cite{jin1996single}. This paper presents the design, experimental tests, and typical results using the diode laser at 852.3 nm of the Cs D2 line and the commercially available 852 nm narrow-band IF from LaserOptik (Germany) as an example. This design can be extended to other wavelengths of IFs combined with cat-eye reflectors for ECDL.

\section{Principle of external-cavity diode laser with dual IFs}
\subsection{External-cavity diode laser with IF}
The diode laser works by the excitation method, using the transition of electrons between the energy bands. The front and rear surfaces of the laser diode form a resonant cavity to provide feedback capability such that the excited radiation photons in the cavity move back and forth many times, forming coherent continuous oscillation, the cavity oscillation of the beam direction and frequency is limited to ensure that the output laser has certain monochromaticity and directivity.

Diode lasers have a wide linewidth owing to their short internal cavity length and low reflectivity at the end surfaces. The principle of the IF-ECDL narrowing laser linewidth is shown in Figure 1. The narrowband IF is usually coated with a multilayer dielectric film, which is an optical element that uses the interference principle to allow a specific spectral range of the laser to pass. The Eq.(1) \cite{yeh1990optical} represents the central wavelength of the transmission peak $\lambda(\theta)$
\begin{equation}
\lambda(\theta) =
 \lambda_0\; \sqrt{1-{\frac{\sin^2\theta}{n_{eff}^2}}}.
\end{equation}
where $\theta$ is the angle of incidence, $\lambda(\theta)$ is the central wavelength of IF transmission at the incidence angle of $\theta$, $n_{eff}$ is the effective refractive index of IF, and $\lambda_0$ is the central wavelength of IF transmission at the incidence angle of 0°.

As the incident angle $\theta$ changes, the central wavelength $\lambda(\theta)$  of the transmission peak follows. Consequently, the laser longitudinal mode can be selected when rotating the IF. The commercially available narrow-band interference filters are designed for an angle of incidence of 6°, the peak wavelength of 852.3 nm, and the peak transmittance of 80\% to 85\%, which is typical of Laser Optik's designs. The main reason for this design is to allow the narrow-band interference filters to operate at the peak wavelength without the reflected beam feedback to the diode laser, avoiding the effects of optical feedback on laser stability.
\begin{figure}[ht!]
\centering\includegraphics[width=6cm]{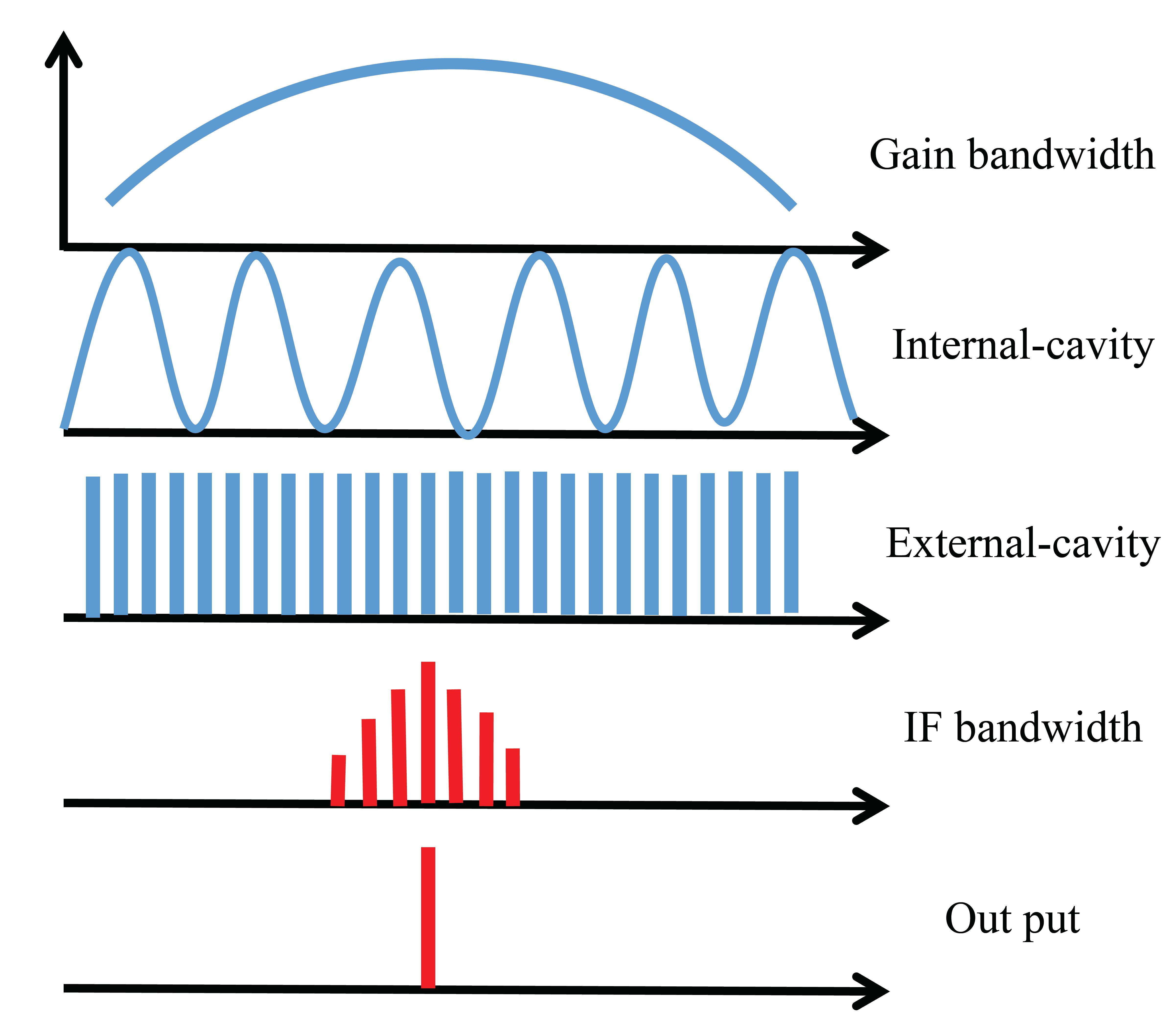}
\caption{IF-ECDL narrowing laser linewidth principle. The output laser is collimated to a parallel beam from the laser diode, oscillates in the external cavity, and is filtered many times by the narrow-band interference filter to obtain narrow linewidth laser.}
\end{figure}
\subsection{Selection and combination of narrow-band IFs}

Narrow-band IFs from the same company and batch may also differ in the peak wavelength of the normal incidence. Therefore, in our experiments, we used a Ti sapphire laser to select the suitable IFs. The wavelength of the output laser from the Ti sapphire laser is changed at the same angle of incidence, and the transmission of the IFs was measured at different wavelengths at that angle of incidence, as shown in Figure 2(a). The transmission spectra of the IFs were obtained by selecting different incident angles and repeating this procedure. The typical transmission spectrum of the IF has a transmittance of approximately 90\%  and a bandwidth of approximately 0.5 nm. Its central wavelength decreased with an increase in the angle of incidence. Based on the transmission spectral data, three IFs were selected from the seven pieces. The central wavelengths of the three selected interference filters (Table. 1 is typical parameters) at different angles of incidence and fitted with Eq.(1) to obtain the effective refractive indices and wavelengths at normal incidence, as shown in Figure 2(b).
\begin{figure}[H]
\centering\includegraphics[width=12.5cm]{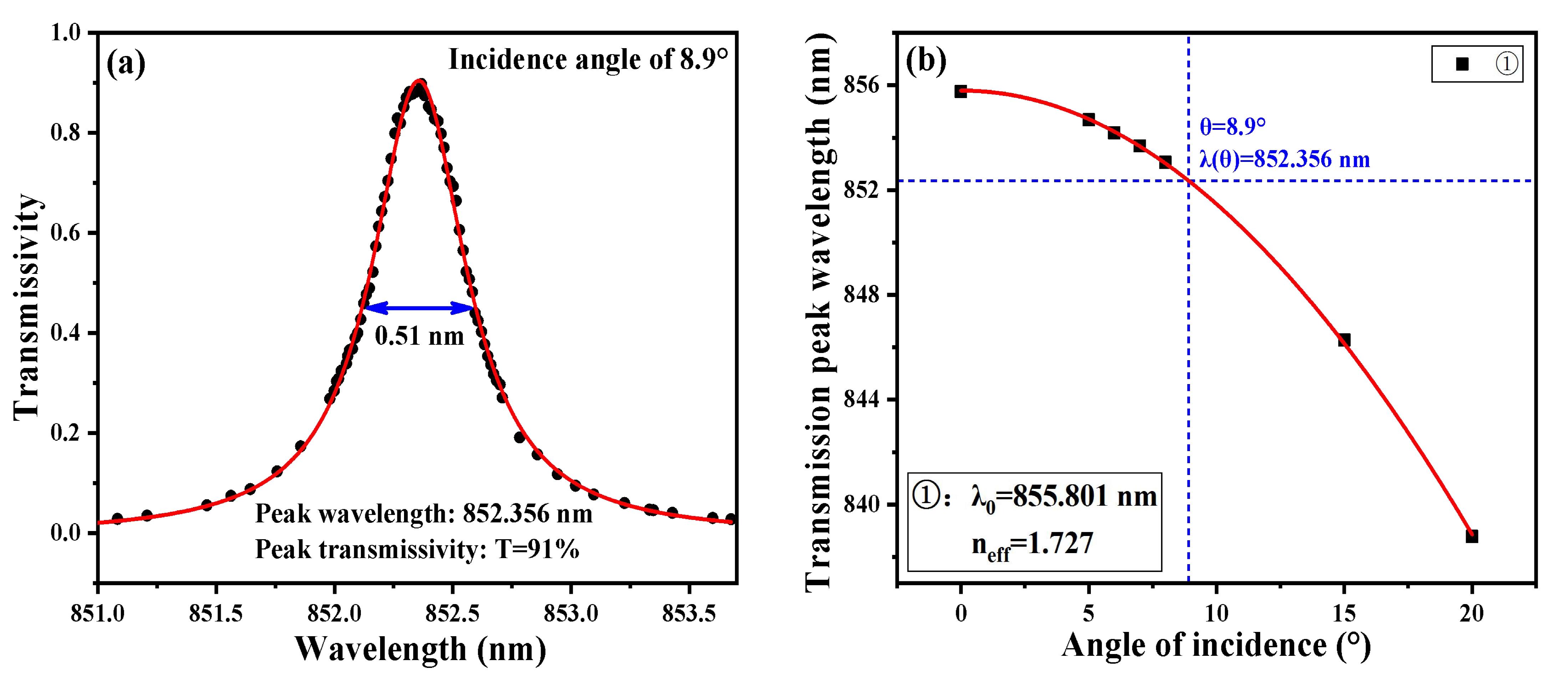}
\caption{Measurement of interference filters parameters. (a) Transmission spectrum of the interference filter at an incidence angle of 8.9°. The wavelength of the Ti sapphire laser is changed, measurements of the transmittance at different wavelengths are performed, and the Lorentz function is used for fitting to obtain the central wavelength and bandwidth of the IF.  Measurements at 9° angle of incidence, correspond to the case of IF in Single-IF-ECDL. (b) Corresponding to different central wavelengths of IFs at different incidence angles are fitted to obtain the effective refractive index. By fitting with Eq.(1), the wavelength of narrow-band IF at normal incidence is 855.801 nm, the parameters of the other IFs are also obtained in this manner.}
\end{figure}

\begin{center}
\begin{table}[H]
\caption{Typical parameters of the interference filters measured in the experiment.}
\centering
\begin{tabular}{cccc}
\br
NO. & Normal incidence wavelength
(nm) & Bandwidth
(nm) & Effective refractive index\\
\mr
\ding{172} & 855.801 & 0.51 & 1.727 \\
\ding{173} & 855.429 & 0.51 & 1.724 \\
\ding{174} & 855.431 & 0.51 & 1.725 \\
\br
\end{tabular}
\end{table}
\end{center}

The output laser linewidth of the IF-ECDL is related to the cavity length of the external cavity and the bandwidth of the IF. Currently, a single IF limits the linewidth of the laser, the bandwidth (full width at half maximum) of the 852 nm narrow-band interference filters that we used in our experiments are designed to be 0.37 nm, but the actual measured bandwidth is ~0.5 nm, which is the smallest bandwidth of this type of commercially available narrow-band interference filters, and due to the processing technology, narrower bandwidth is very difficult to achieve. Therefore, based on a single IF, two IFs were combined to further narrow the laser linewidth at the expense of a small amount of transmittance. There are two ways to combine the IFs. First, the two IFs can be placed in parallel. therein, if two IFs with the same peak wavelength and bandwidth are used, the combined bandwidth is not narrowed, and the transmittance is lowered, so we wished to narrow the bandwidth of the combined IFs and select the difference in peak transmittance wavelengths at normal incidence. In this case, the bandwidth of the combination is limited by the central wavelength difference between the two chosen IFs. In the experiment, we adjusted the combined centre wavelength to the common wavelength used in cesium atom-related experiments, such as 852.3 nm for Cs atoms related experiments. We selected two IFs (\ding{172} and \ding{173} in Table.1) with the same bandwidth and different central wavelengths placed parallel at an incidence angle $\theta$ of 8.7°. The combined simulation is shown in Figure 3(a). 

\begin{figure}[ht!]
\centering\includegraphics[width=12.5cm]{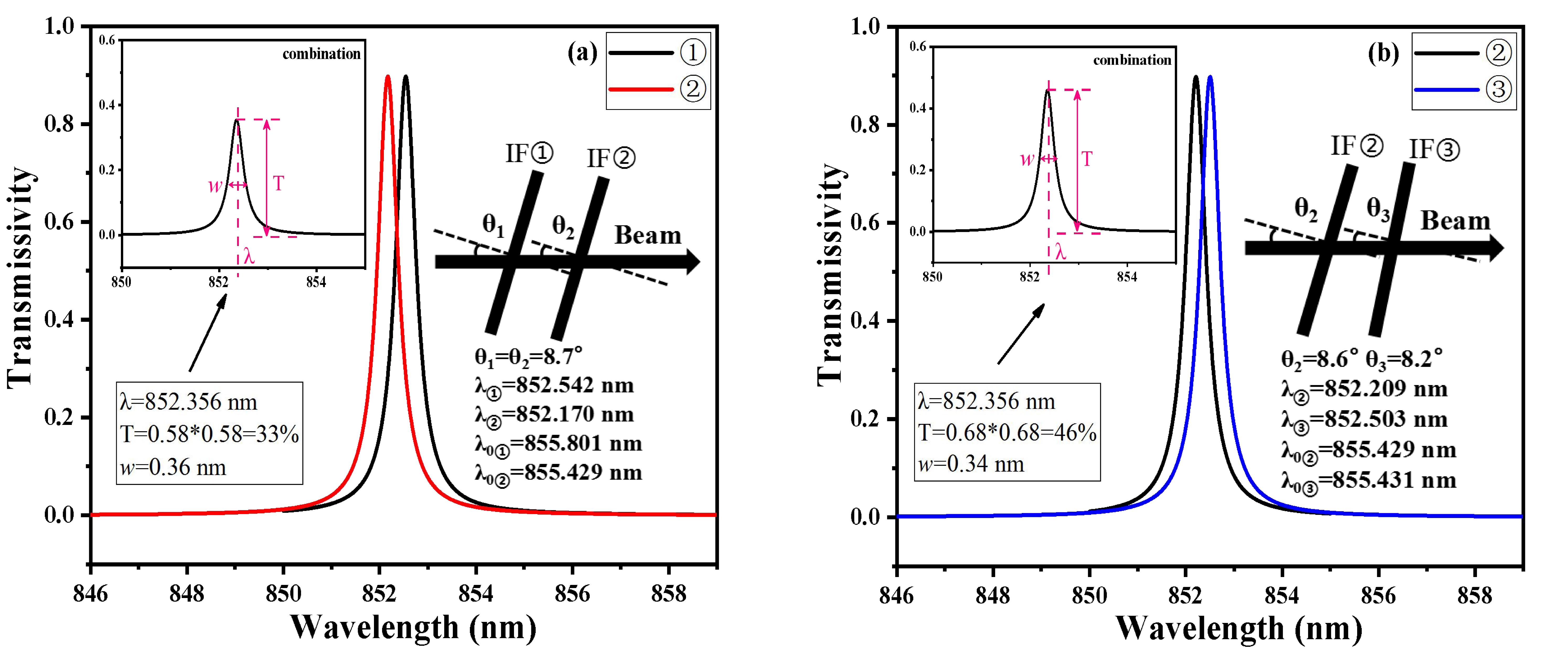}
\caption{The IFs combinations according to scheme 1 and scheme 2, with a central wavelength of 852.356 nm, corresponding to the Cs 6S$_{1/2}$(F=4)-6P$_{3/2}$(F$^\prime$=3,4,5) transitions. (a) According to the scheme 1, the two IFs are combined with the same bandwidth but different central wavelengths. The normal incident central wavelengths of the two IFs are 855.801 and 855.429 nm, respectively. The peak transmissivity and FWHM are 90\% and 0.51 nm, respectively. They are placed in parallel at an incidence angle of 8.7°. The left inset shows that the transmissivity after the combination is 33\% and the bandwidth is 0.36 nm, respectively. The right inset shows that the parallel combination of dual IFs ; (b) According to the scheme 2, the two IFs are combined with the same bandwidth and central wavelengths. The normal incidence central wavelengths of the two IFs are 855.429 nm and 855.431 nm, respectively. The peak transmissivity and FWHM are also 90\% and 0.51 nm, respectively. They are placed at a dihedral angle of 0.4°. The left inset shows that the transmissivity after combination is 46\% and the bandwidth is 0.34 nm, respectively. The right inset shows that the dihedral angle combination of dual IFs.}
\end{figure}
If the peak wavelength and bandwidth are the same at the normal incidence of the two IFs, then an optimal dihedral angle combination is required to achieve a narrower bandwidth. This is the second method; a simulation is illustrated in Figure 3(b). Where two IFs with approximately the same bandwidth and center wavelength (\ding{173} and \ding{174} in Table.1) are selected, and their respective incidence angles are adjusted such that a dihedral angle is formed between the two IFs. The transmission spectrum of the IF is Lorentzian; therefore the combination of two IFs is effective to the combination of two Lorentzian transmission spectra with different centers, and the bandwidth will be smaller when the difference between the two Lorentzian peak center wavelengths is within a certain range, and the transmittance will also be smaller. Therefore, the optimal value of the combined effect of transmittance and bandwidth should be considered when combining the two IFs. The Lorentz function is expressed as

\begin{equation}
y_1 = y_0+\frac{2A}{\pi}\cdot \frac{w}{4x^2+w^2}.
\end{equation}
\begin{equation}
y_2 = y_0+\frac{2A}{\pi}\cdot \frac{w}{4(x-\bigtriangleup \lambda)^2+w^2}.
\end{equation}
\begin{equation}
y_m = y_1-y_2=\frac{2Aw}{\pi}\cdot \frac{4(\bigtriangleup\lambda)^2-8x(\bigtriangleup\lambda)}{(4x^2+w^2)\left[4(x-\bigtriangleup\lambda)^2+w^2\right]}.
\end{equation}
where A is the area of the transmission peak, $\bigtriangleup\lambda$ =$\left | \lambda_2-\lambda_1\right |$, and $\textit{w}$ is the full width at half maximum of the Lorentzian shape, which is the bandwidth of the IF here.

The difference between the two Lorentz functions with different central wavelengths was used to obtain Eq. (4). The zeros of this function can be used to describe the combined effects of bandwidth and transmittance of the dual IFs. Derive the zeros of this function as
\begin{align}
\frac{d y_{m}}{d x} =& \frac{2 A w}{\pi} \cdot\left\{\frac{-8 x(\Delta \lambda)}{\left(4 x^{2}+w^{2}\right)\left[4(x-\Delta \lambda)^{2}+w^{2}\right]}\right\}-\frac{2 A w}{\pi} \cdot\left\{\frac{8 x\left[4(\Delta \lambda)^{2}-8 x(\Delta \lambda)\right]}{\left(4 x^{2}+w^{2}\right)^{2}\left[4(x-\Delta \lambda)^{2}+w^{2}\right]}\right\} \\ \nonumber
&-\frac{2 A w}{\pi} \cdot\left\{\frac{8(x-\Delta \lambda)\left[4(\Delta \lambda)^{2}-8 x(\Delta \lambda)\right]}{\left(4 x^{2}+w^{2}\right)\left[4(x-\Delta \lambda)^{2}+w^{2}\right]^{2}}\right\}
\end{align}
In the experiment, we needed the optimal value of the combined effect of the bandwidth and transmittance after the two IFs, so we took the extreme value of the zeros of this function. When $\frac{d y_{m}}{dxd(\Delta \lambda)}=0$, can obtain $\Delta \lambda=\frac{\sqrt{3}}{3}\textit{w}$. Further, when \textit{w}=0.51 nm, $\Delta \lambda$=0.294 nm is obtained, and $\Delta \theta$=0.4° is obtained via Eq.(1). This is the second method combining two IFs. The simulation results of Figure 3(a) and (b), show that the two IFs with the special dihedral angle combination had greater transmittance and narrower bandwidth than the IFs in the parallel combination; in the second method, it is easier to obtain a narrow FWHM, which is not limited by the wavelength difference of the two IFs, in this case, the combined IFs transmittance is higher, the insertion loss is lower, and the laser output power is higher.

\section{Experimental setup}
\subsection{Construction of the IF-ECDL}
The structure of the IF-ECDL is shown in Figure 4(a) and Figure 4(b). The scheme adopted in this study is shown in Figure 4(b), the physical diagram of which is shown in Figure 5(a). The advantage of this solution is that it avoids the deterioration of the beam quality caused by the mechanical non-coaxial nature of the two cat-eye lenses as well as cost savings, and the combination of a half-wave plate and a polarization beam splitting prism (PBS) facilitates a wide range of tuned feedback.
\begin{figure}[ht!]
\centering\includegraphics[width=7cm]{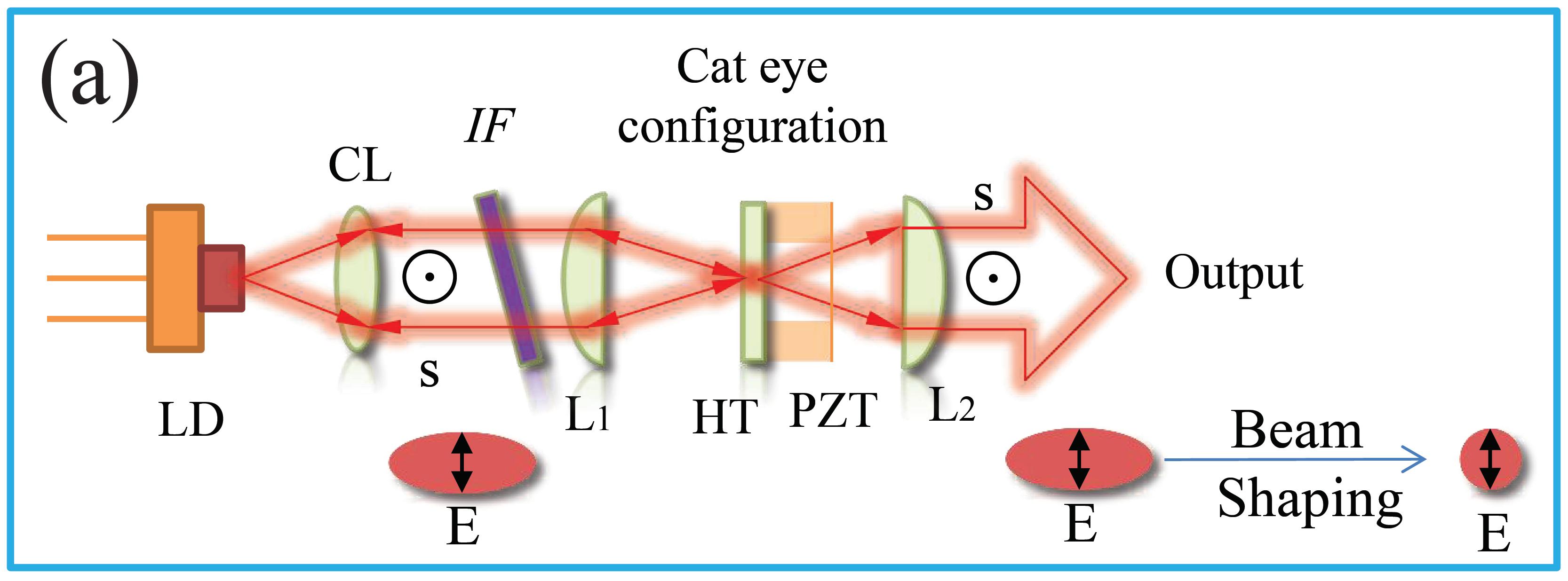}\\
\includegraphics[width=7cm]{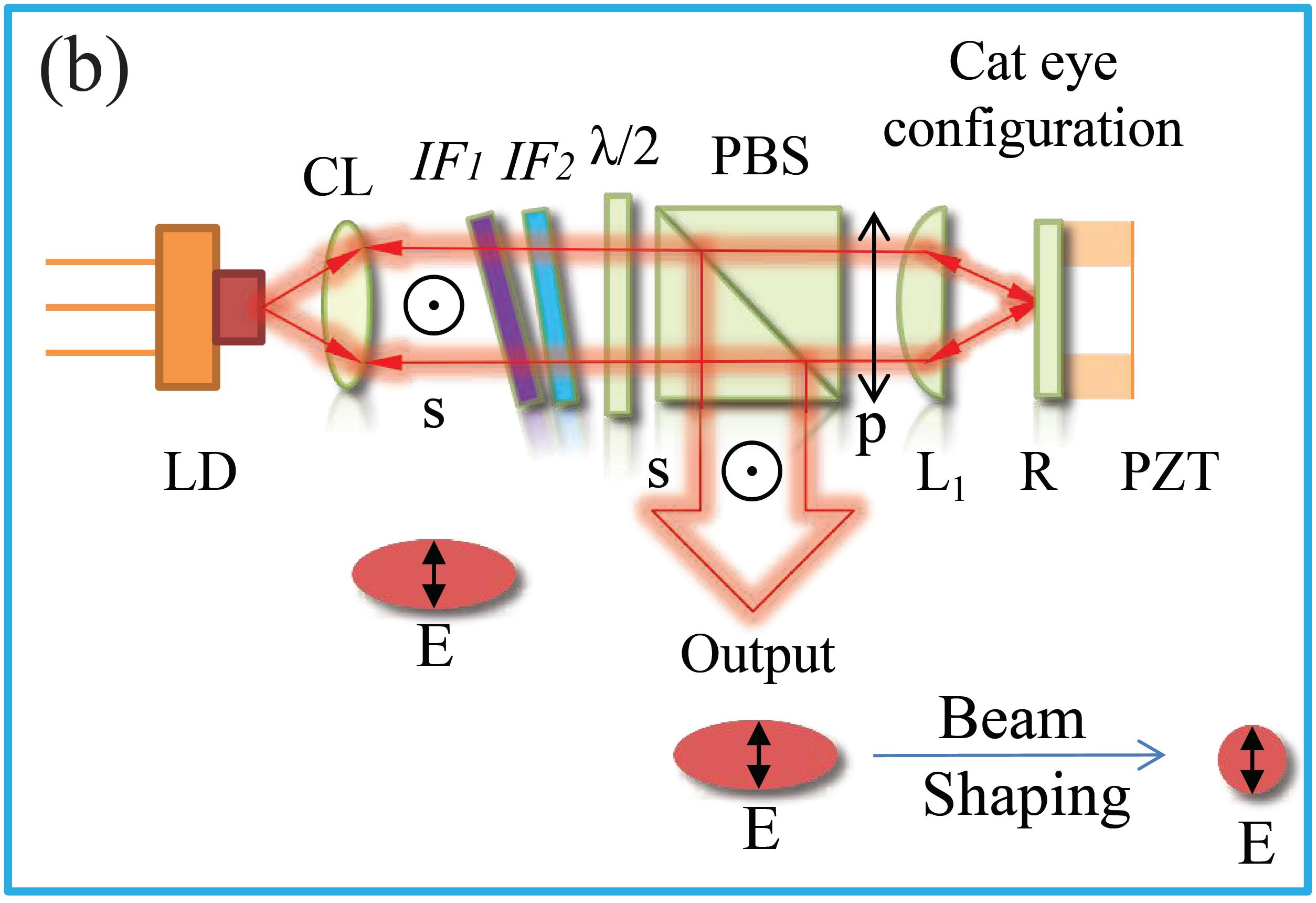}
\caption{Schematic of the IF-ECDL structure. (a) Conventional IF-ECDL, which uses one IF in the external cavity, the feedback of the external cavity is not adjustable unless the reflector is replaced with a different reflectivity. (b) Modified Dual-IF-ECDL. LD: Laser diode; CL: Aspherical collimating lens; IF: Narrow-band interference filter; HT: High-transimissivity mirror; $\lambda/2$: Half-wave plate; PBS: Polarizing beam splitter cube; L1 and L2: cat-eye lens; R: High-reflectivity mirror; PZT: Piezoelectric ceramic; s: s polarization; p: p polarization.}
\end{figure}

The anti-reflection coating 852 nm laser diode in a 9-mm-can-packaged was used as the light source in the IF-ECDL, and the laser diode was rotated such that the output laser was an elliptical beam with a long axis parallel to the horizontal, at which point the output laser was s-polarized. After the output laser was collimated by an aspherical lens with f=8 mm, the longitudinal mode was selected by a narrow-band IF, the beam was passed through the half-wave plate and PBS, and the p-polarized beam was focused by the cat-eye lens. The lens focused on the reflector, and the angle of the reflector was adjusted such that the beam returned along the original path, forming a cat-eye effect \cite{baillard2006interference}. Thus, the P-polarized beam was reflected back to the laser diode, and the PBS reflected the S-polarized beam as the output laser. Further, the output laser was changed to a near-circular beam by an Anamorphic Prism pair. The feedback can be controlled by adjusting the half-wave plate, which facilitates the investigation of the laser feedback in relation to the laser linewidth and tuning range.  Figure 5(b) shows the controller used by the IF-ECDL because of the need for good temperature control of the diode laser. Brass with good thermal conductivity and easy machining was chosen as the base. To improve the stability of the laser, we mounted the laser in a conductive aluminium enclosure, forming a Faraday-cage, which shields the electromagnetic radiation from the outside, as shown in Figure 5(c). The laser linewidth of the IF-ECDL depends on the cavity length and the bandwidth of the narrow-band IF. The longer the cavity length, the narrower the laser linewidth. However, the smaller the longitudinal mode spacing, the more easily the modes change when scanning the laser frequency.

\begin{figure}[H]
\centering\includegraphics[width=6cm]{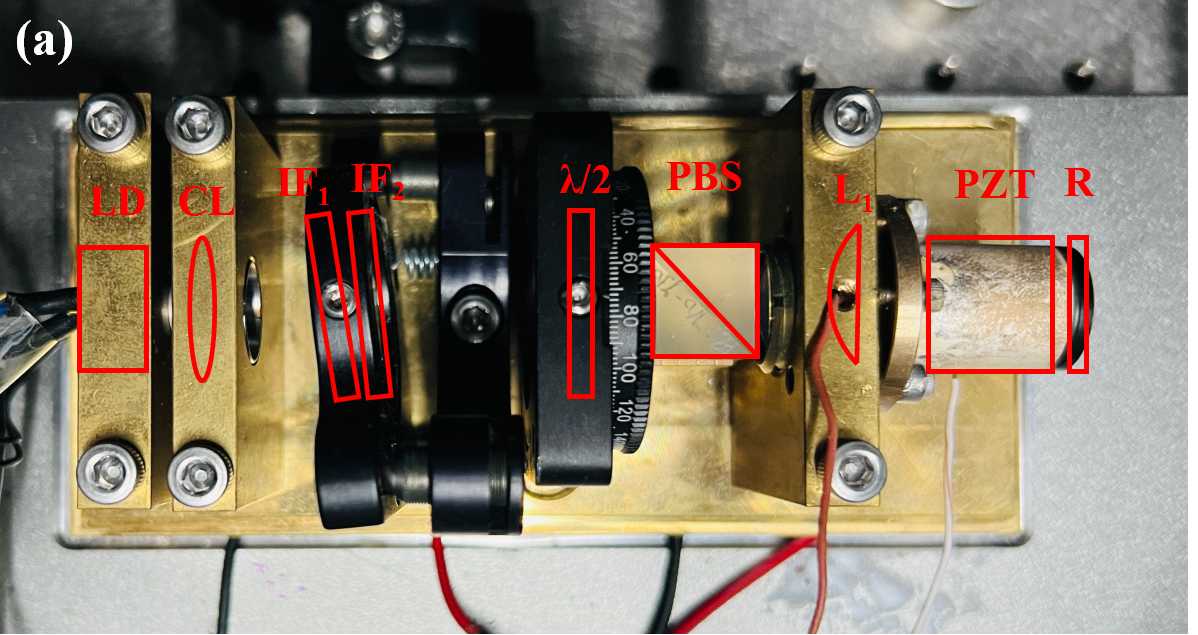}
\includegraphics[width=6.5cm]{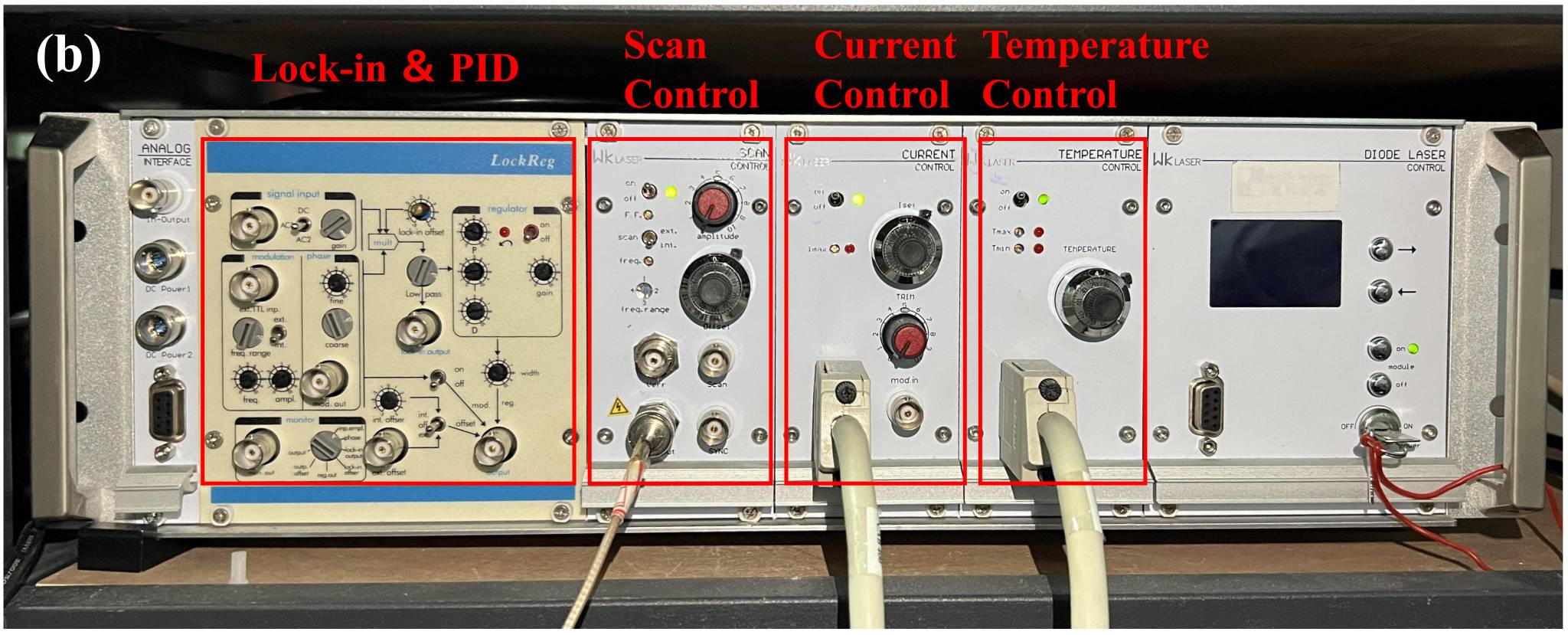}\includegraphics[width=3.5cm]{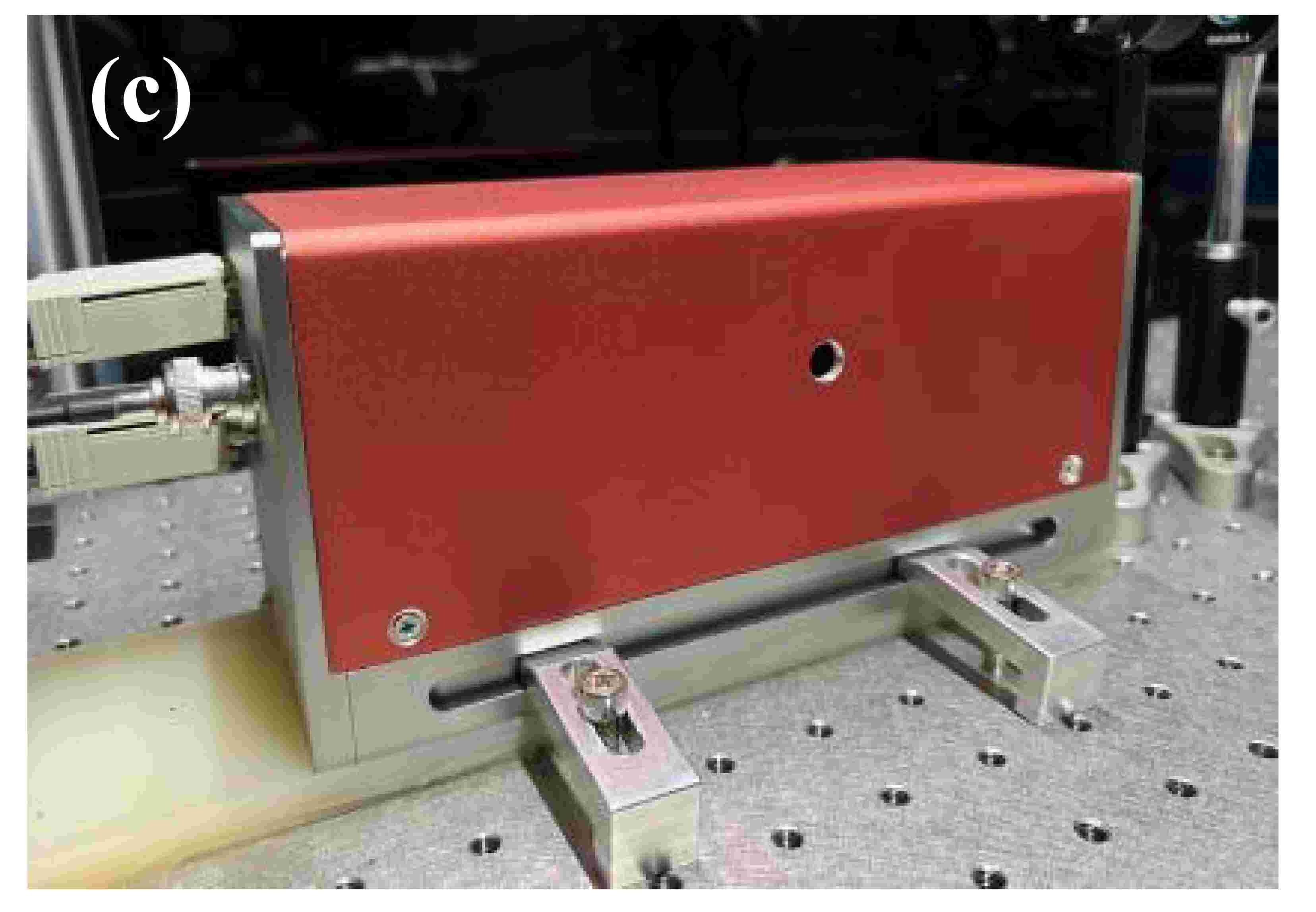}
\caption{ Physical diagram of the Dual-IF-ECDL system. (a) Photo of the Modified Dual-IF-ECDL. The base is made of brass because of its good thermal conductivity, which is connected to the temperature controller by TEC elements and thermistor for precise temperature control, and an organic glass cover (removed) facilitates temperature control; (b) Photo of the Dual-IF-ECDL controller. It includes a temperature control module, a current control module, a scanning module, and a lock-in \& PID module; (c) Photo of the laser head, the aluminum enclosure contains all the components shown in (a) and the temperature controlled brass base, which forms a Faraday-cage to shield the electromagnetic radiation from the outside.}
\end{figure}

\subsection{Three scenarios for IF-ECDL}
Based on the number of narrow-band interference filters and the combination method, we have designed the IF-ECDL for three cases: Single-IF-ECDL (852 nm) in the single narrow-band IF case, and two combinations in the two narrow-band IFs cases: one is the Dual-IF-ECDL-$\#$1 combination in Figure 3(a), and the Dual-IF-ECDL-$\#$2 combination in Figure 3(b). After comparing the experimental measurements, the results in Figure 10 show that the Dual-IF-ECDL-$\#$2 outputs narrower laser linewidth.

\section{Experimental results}
\subsection{IF-ECDL output VS the injection current and beam quality}

By controlling the laser temperature at 22.5 °C and gradually increasing the current to 130 mA in the bare LD case, the Dual-IF-ECDL-\#2 output power was approximately 22.3 mW. With the addition of an external cavity to the laser, and the feedback was optimized by adjusting the focal position of the cat-eye lens and the angle of the reflector. Optimal feedback is achieved when the current threshold is reduced to a minimum value \cite{ye2004tunable,pan2018broad}. When the front surface of the LD forms an external cavity with the reflector, the beam is reflected many times in the external cavity, enabling the gain core area of the LD to be fully utilized. This reduces the current threshold of the IF-ECDL, and increases the output power. Above the current threshold, the output power of the IF-ECDL is linearly related to the current above the threshold, as shown in Figure 6.

\begin{figure}[H]
\centering\includegraphics[width=7cm]{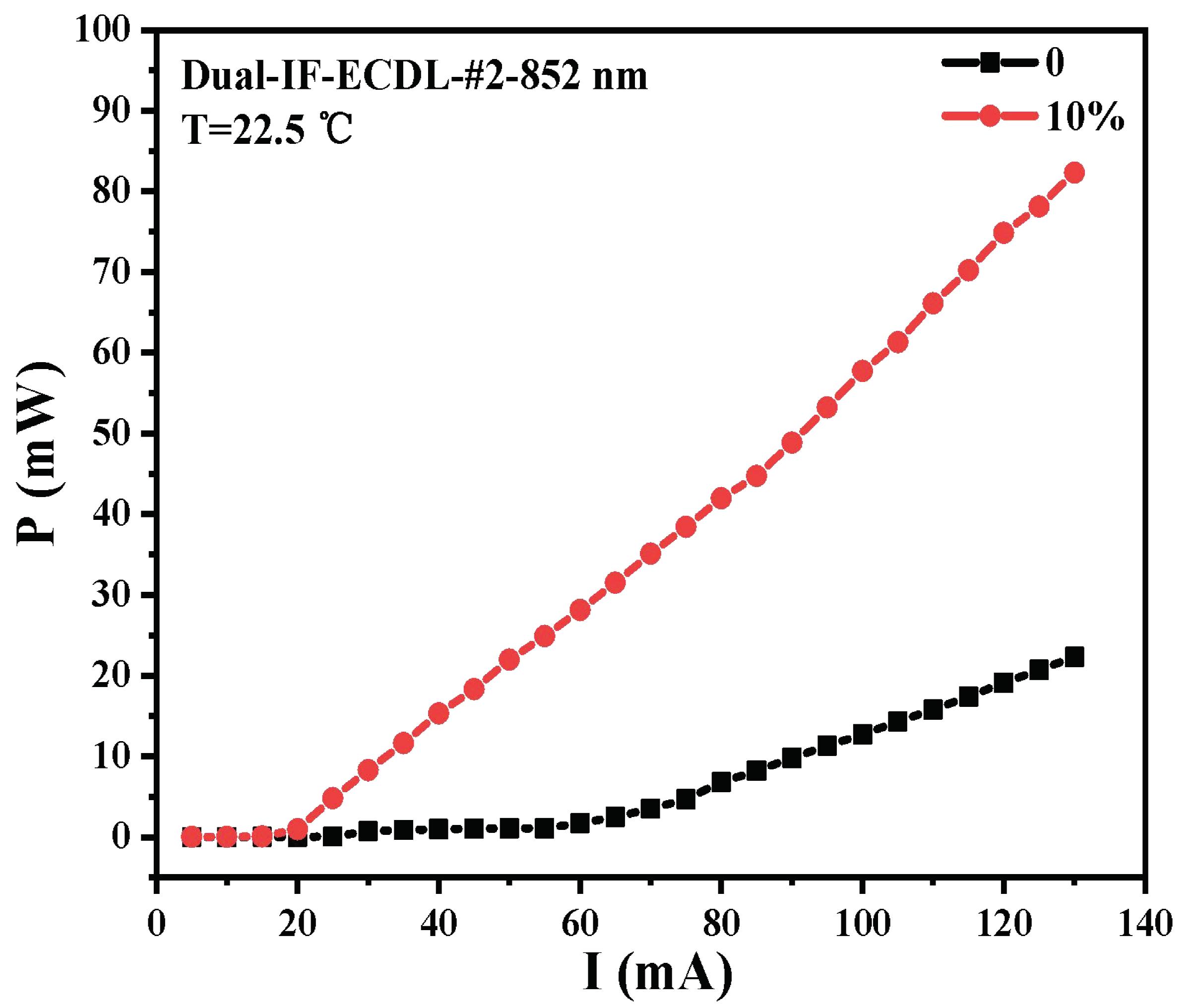}
\caption{ Dual-IF-ECDL-\#2-852 nm output VS the injection current. Above the threshold, the current of the IF-ECDL is linearly related to the output power, At 10\% feedback, the current threshold is reduced from 55 to 20 mA and the slope is increased from 0.27 to 0.73.}
\end{figure}
Next, we measured the beam of the IF-ECDL, which outputs an elliptical beam that becomes a near-circular beam after passing through a shaping prism. The obtained beam quality $M^2$ of the Dual-IF-ECDL is measured by a laser beam profiler, and the measured $M^2$ is ${M^2}_{horizontal}$=1.15, ${M^2}_{vertical}$=1.18, as shown in Figure 7.
\begin{figure}[H]
\centering\includegraphics[width=6.6cm]{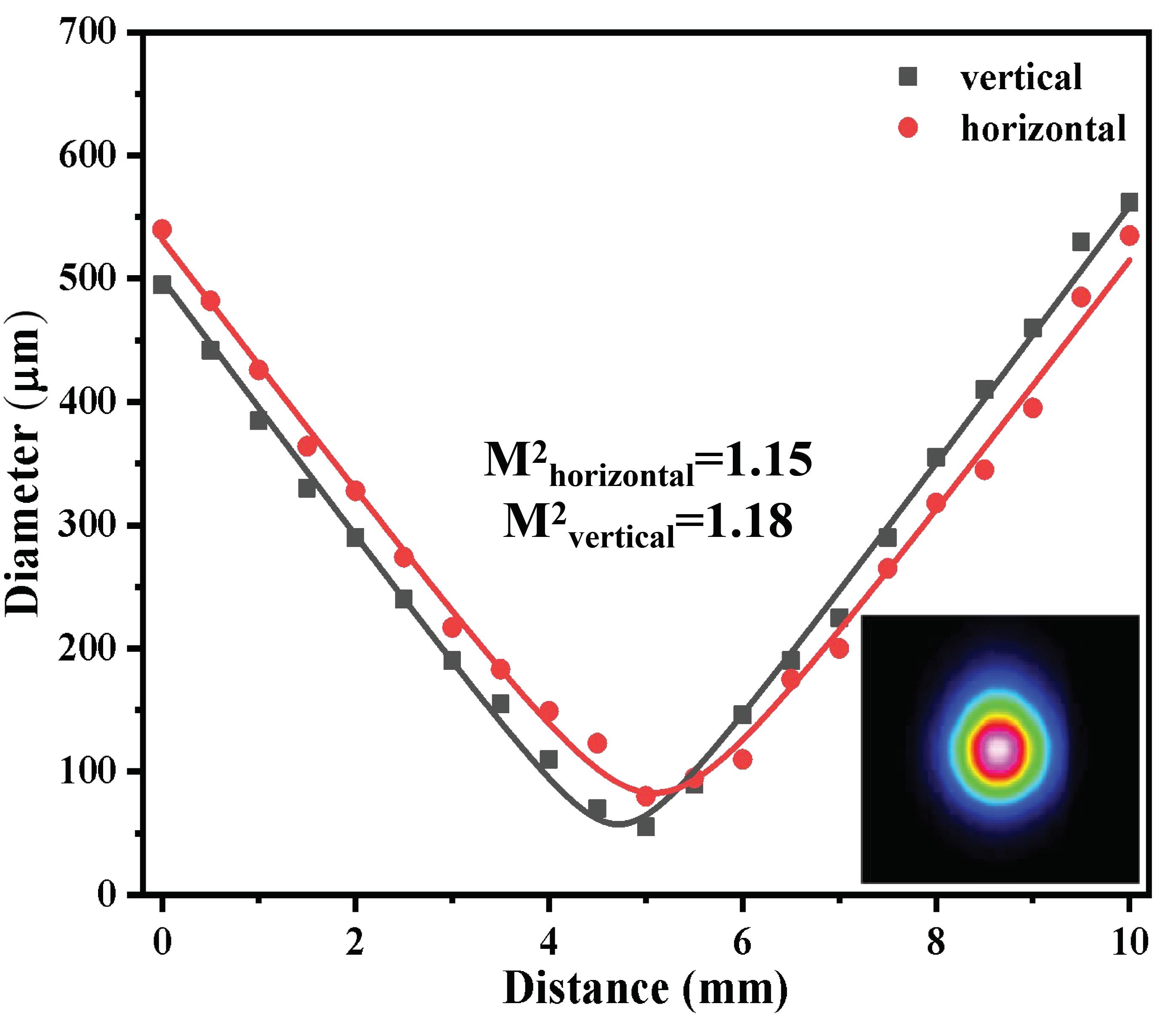}
\caption{The Dual-IF-ECDL-$\#$2-852 nm collimated and shaped spot intensity distribution, and measurement of $M^2$. The ${M^2}_{horizontal}$=1.15, ${M^2}_{vertical}$=1.18.}
\end{figure}

\subsection{Saturation absorption spectroscopy and IF-ECDL frequency stabilization}

In the experiments related to atoms, the state of the atoms was manipulated with a laser. Atoms have a stable and precise energy-level structure; thus, the proposed laser must have a precise and stable frequency. However, instabilities in laser temperature, drive current and mechanical vibrations cause the laser frequency to drift. To obtain a frequency-stable laser, it must be locked to a stable frequency reference standard such as atomic spectra. Saturation absorption spectroscopy is a typical method for locking the laser frequency. In our experiments, the Dual-IF-ECDL output laser with a continuous frequency tuning range greater than 9.2 GHz with simultaneous scanning of the LD injection current and PZT of the external cavity. This is reflected that two saturation absorption spectra of Cs (F=3) and (F=4) could be scanned corresponding to laser wavelengths of 852.335 nm and 852.356 nm, respectively, with the frequency difference of approximately 9.2 GHz. The saturation absorption spectra obtained are shown in Figure 8(a) and (b).
\begin{figure}[ht!]
\centering\includegraphics[width=13cm]{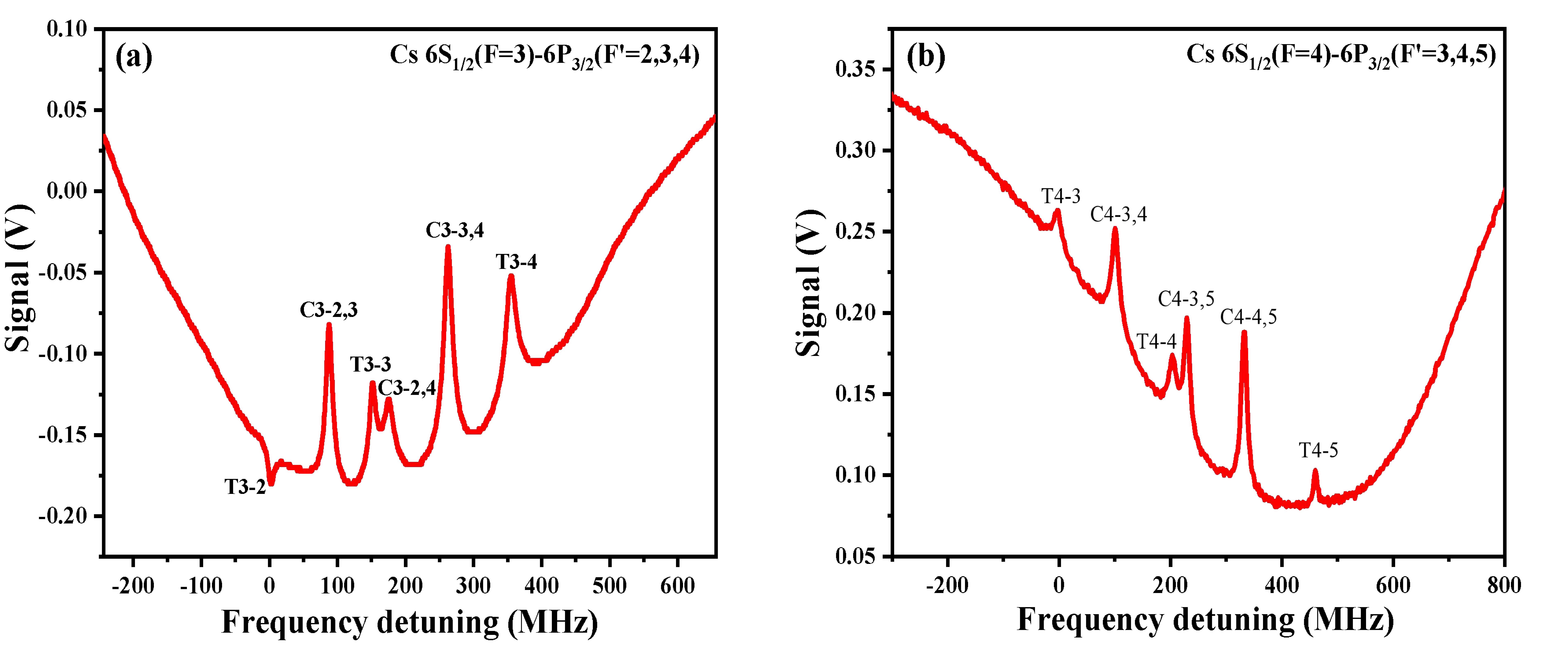}
\caption{D2 line saturation absorption spectra of Cs (F=3) and (F=4). The two corresponding laser wavelengths are 852.335 and 852.356 nm respectively, with the frequency difference of approximately 9.2 GHz. (a) saturation absorption spectrum of Cs 6S$_{1/2}$(F=3)-6P$_{3/2}$(F$^\prime$=2,3,4); (b) saturation absorption spectrum of Cs 6S$_{1/2}$ (F=4)-6$P_{3/2}$(F$^\prime$=3,4,5).}
\end{figure}

Frequency stability is an important characteristic of diode lasers. We experimentally recorded the frequency drift of the Dual-IF-ECDL-\#2 free-running and frequency-locked states at the hyperfine transition of the Cs atoms. The atoms at room temperature have strong thermal motion with Boltzmann distribution, which leads to the Doppler broadening of the absorption spectrum, the laser is absorbed over a wide frequency range. We take the absorption spectrum through the lock-in amplifier to obtain a first-order differential signal. This method was used to assess the frequency drift when the laser was free-running. We used the saturated absorption spectrum to stabilize the laser frequency, which was modulated in the experiment. The saturated absorption spectrum after frequency modulation contained the first-order differential signal (error sign), which was obtained through subsequent filtering and amplification and the feedback circuit was used to control the laser frequency. These two cases were recorded for 30 minutes, and the results showed that the frequency drift was reduced by approximately 40 times, as shown in Figure 9.

\begin{figure}[H]
\centering\includegraphics[width=13cm]{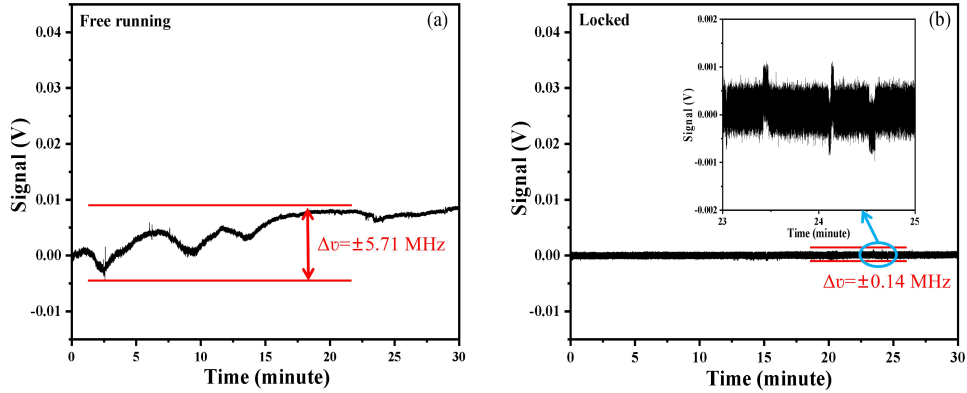}
\caption{Measurement of Dual-IF-ECDL-\#2 frequency stability. (a) Frequency drift of Dual-IF-ECDL-\#2 under free-running conditions, $\Delta\nu$=±5.71 MHz; (b) The frequency drift when the laser frequency is stabilized in the saturated absorption spectrum of Cs 6$S_{1/2}$(F=4)-6$P_{3/2}$(F$^\prime$=5), $\Delta\nu$=±0.14 MHz. These two cases were recorded for 30 minutes, and the results show that the frequency drift was reduced 40 times.}
\end{figure}
\subsection{Measurement of the IF-ECDL laser linewidth}

We measured the laser linewidth of the IF-ECDL using an optical F-P cavity  \cite{Li1991}. When using the F-P cavity to measure the laser linewidth, if the cavity linewidth is smaller than the laser linewidth, the transmitted linewidth is approximately equal to the laser linewidth; when the cavity linewidth is approximately the same as the laser linewidth, the transmitted linewidth is the combined linewidth of the two; when the cavity linewidth is larger than the laser linewidth, the transmitted linewidth is approximately equal to the cavity linewidth. So, this method requires that the linewidth of the F-P cavity is less than the laser linewidth to be measured. When the FSR of the cavity is certain, the more fineness, the smaller of the cavity linewidth. Therefore, we used a highfinesse F-P cavity for the measurement, as shown in Figure 10(a). The output laser of the IF-ECDL entered the electro-optic phase modulator (EOPM) and then the high-finesse F-P cavity, which had a length of 100 mm and a cavity linewidth of approximately 15 kHz. The transmitted signal of the cavity was observed with a photodetector. The linewidth was measured by using the signal generator (DS345) to add sine-wave modulation at 1 MHz to the EOPM. When the crystal in the EOPM is modulated by an applied electric field, the refractive index changes continuously with the modulating signal, producing a specific frequency difference in the laser, which is the frequency scale. We used this 1 MHz frequency scale to measure the FWHM of the peak transmission signal as the laser linewidth.

We measured and compared the linewidths of the IF-ECDL for three cases: the single IF, the two IFs placed in parallel, and the two IFs placed at a dihedral angle. The measured results are shown in Figure 10(b), (c), and (d), respectively.

Because the laser linewidth of the IF-ECDL is related to the bandwidth of the IF, the combination has a narrower bandwidth than that of the single IF. The comparison of the results in Figure 10(b) and (c) shows that the two IF combinations allowed further narrowing of the laser linewidth. From the results in Figure 10(c) and (d), it can be concluded that when the two filters were placed at a specific dihedral angle, the laser linewidth was narrower and the transmission was greater than when the two filters were placed in parallel. This is in general agreement with the previous simulation results.
\begin{figure}[ht!]
\centering\includegraphics[width=12.5cm]{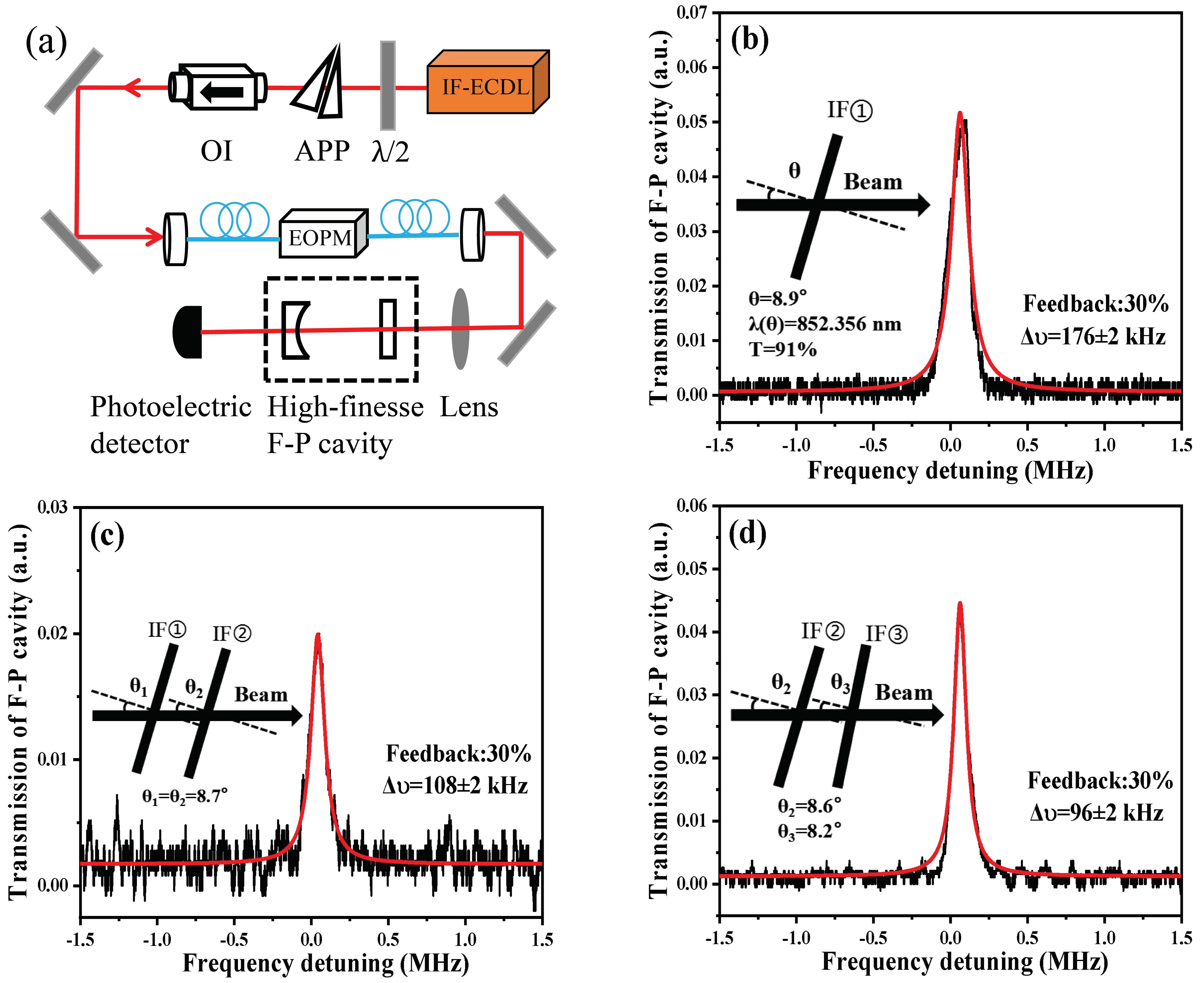}
\caption{Laser linewidth measurement results of IF-ECDL. (a) The output laser linewidth measurement device. The IF-ECDL output laser through the electro-optic phase modulator (EOPM) and an RF modulation signal is applied. The laser linewidth is measured by calibrating the 1MHz sidebands of the high-finesse F–P cavity transmission peak to obtain a frequency scale; (b) Laser linewidth in the single IF case $\Delta\nu$=176±2 kHz. The inset shows the incidence angle of the single IF; (c) Laser linewidth with two IFs placed in parallel $\Delta\nu$=108±2 kHz. The inset shows the parallel combination of dual IFs; (d) Laser linewidth with two IFs placed at a dihedral angle $\Delta\nu$=96±2 kHz. The inset shows the dihedral angle combination of dual IFs. The errors in the measurement result from vibrations caused by noise in the environment or outgassing in the vacuum of the high finesse F-P cavity.}
\end{figure}

To investigate whether the laser linewidth of the IF-ECDL is related to the ratio of the feedback-to-output, the half-wave plate in the IF-ECDL was rotated and the feedback-to-output ratio was varied. Measurements were performed at 10, 20, 30, 40, and 50\% feedback, as shown in Figure 11.

When the other conditions were equal, the laser linewidth varied more markedly from 267 to 134 kHz with an increase in the ratio of the IF-ECDL feedback to the output. As the feedback increased, the laser linewidth in Figure 11(f) exhibited a monotonic decrease. In addition, white noise could be coupled from the RF bias-T module into the DC drive current of the IF-ECDL, with a bandwidth of approximately 100 MHz. Different intensities of the white noise signal have different broadening on the laser linewidth, thus enabling manipulation of the laser linewidth \cite{lu2022analyzing}. The commercial external cavity diode lasers currently used in experiments, whether IFs or grating external-cavity diode lasers, want to change the laser linewidth, you need to replace the interference filters with different bandwidths or grating with different diffraction efficiency.

\begin{figure}[ht!]
\centering\includegraphics[width=6.3cm]{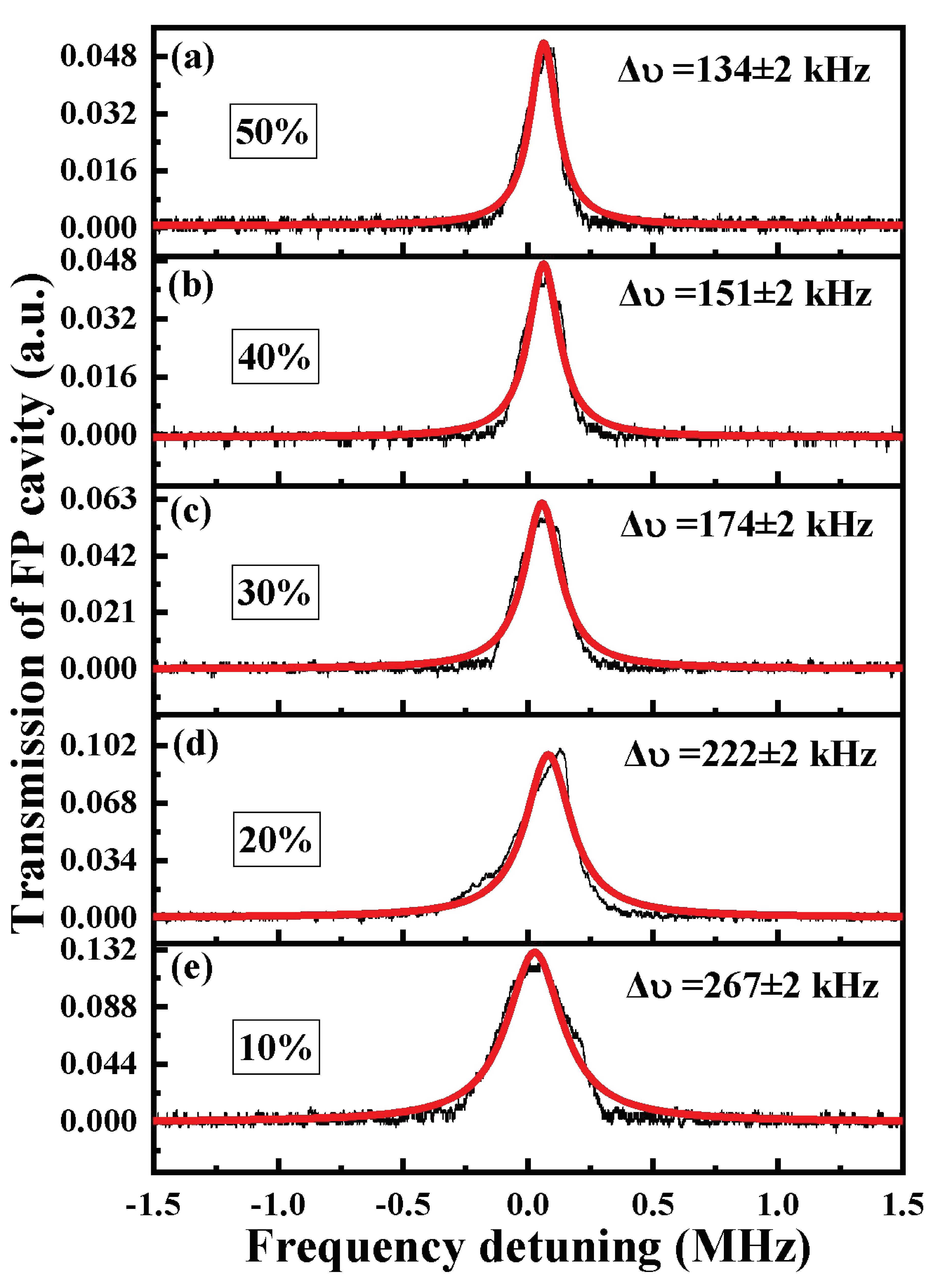}\includegraphics[width=5.5cm]{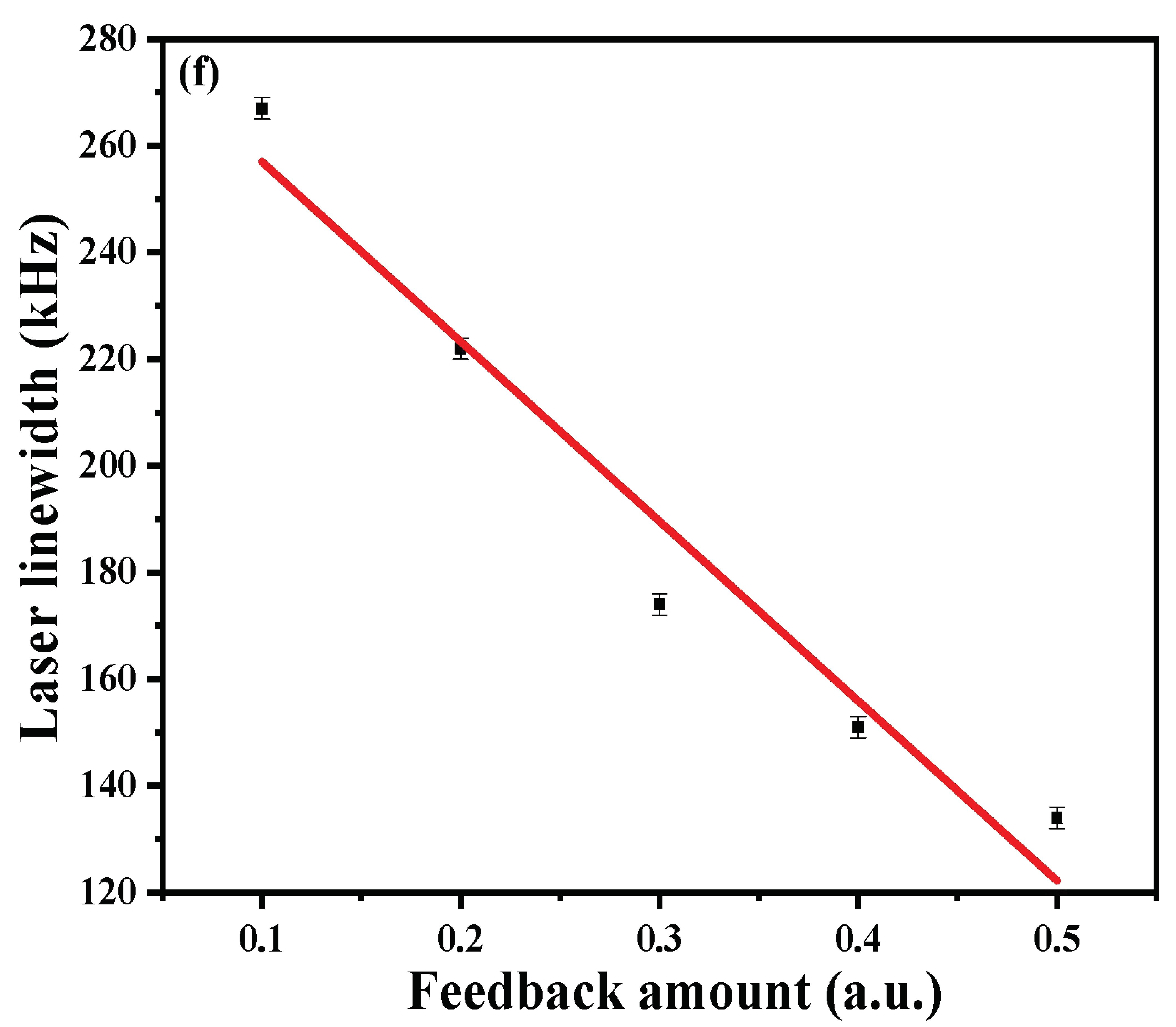}
\caption{Laser linewidth measurement by varying the feedback. (a)-(e) Change the feedback from 10-50\%, IF-ECDL laser linewidth 267±2 kHz-134±2 kHz (f) Variation of laser linewidth with feedback. Error bars are obtained from the Lorentzian fit.}
\end{figure}

\subsection{Measurement of the IF-ECDL frequency tuning range}

\begin{figure}[ht!]
\centering\includegraphics[width=6.7cm]{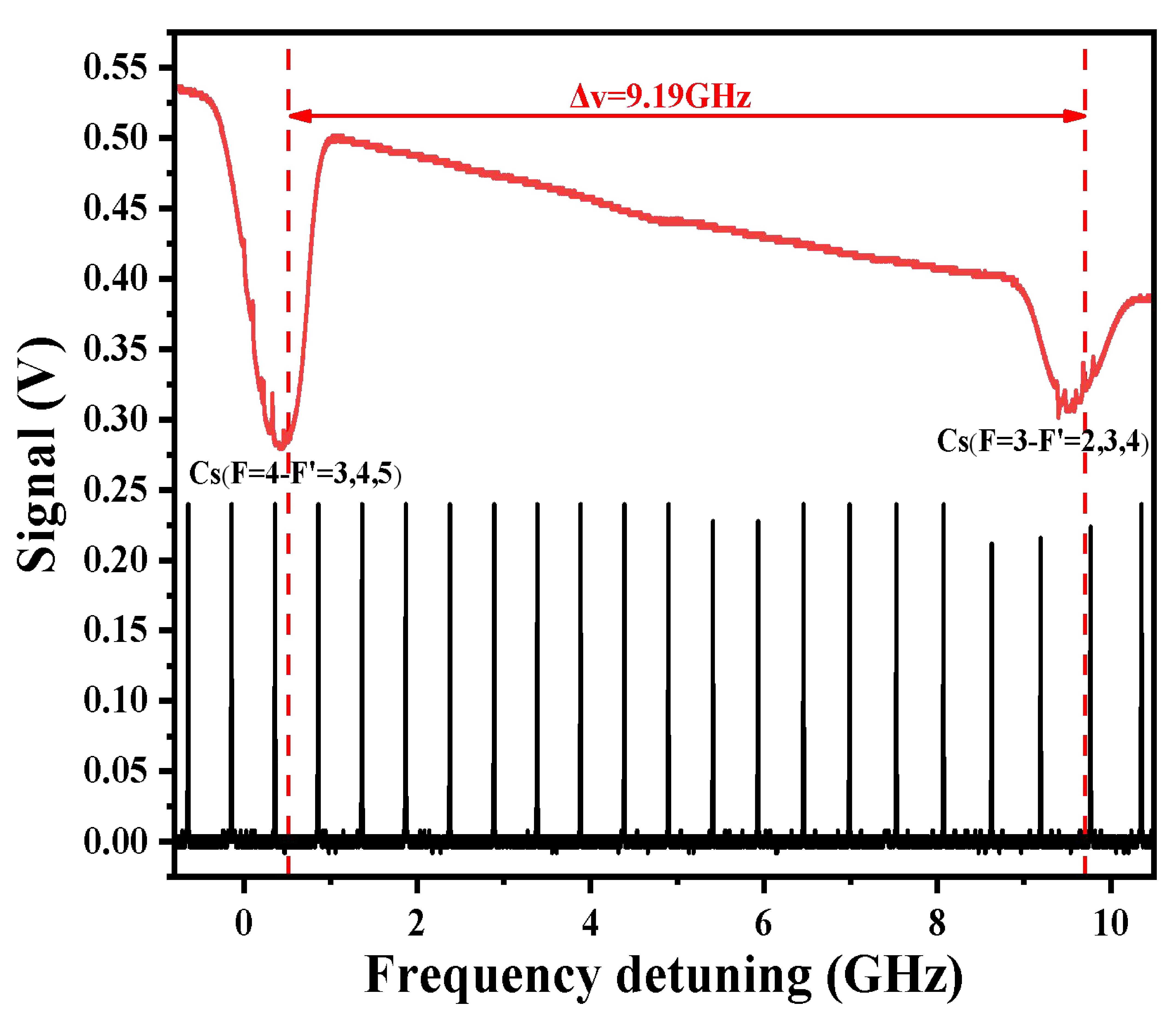}
\caption{F-P cavity calibration laser tuning range corresponding to sweep through the 6S1/2 (F=3)-6P3/2 (F$^\prime$=2,3,4) saturation absorption spectrum and the 6S1/2 (F=4)-6P3/2 (F$^\prime$=3,4,5) saturation absorption spectrum of Cs atoms, spanning the frequency of approximately 9.19 GHz. The calibration of frequency detuning using the 6S1/2 (F=4)-6P3/2 (F$^\prime$=3) saturation absorption spectrum of Cs atoms as reference.}
\end{figure}
The wavelength tuning range of a laser is an important parameter of the ECDL. To render the wavelength tuning range of the IF-ECDL as large as possible, experimentally, we used plated anti-reflection film LD to weaken the mode competition between the internal and external cavity. Further, a cat-eye lens in front of the PZT and the reflector can be used to weaken the instability caused by the non-linearity of the elongation during the PZT scan. In our experiments, we used an F-P cavity to calibrate the wavelength tuning range of the IF-ECDL. The output laser was injected into the F-P cavity, and the PZT was scanned to obtain the spectrum shown in Figure 12. The F-P cavity used had a length of 150 mm and a free spectral range (FSR) of 500 MHz. The laser frequency was scanned to sweep through 20 FSR corresponding to the two saturation absorption spectra of Cs atoms sweeping through approximately 9.19 GHz.

Adjusting the half-wave plate in the IF-ECDL allows more laser to be feedback into the gain core of the laser diode. The greater the feedback the lower the threshold gain of the laser diode and the easier it is to reach the threshold condition, allowing a greater range of wavelength tuning. In our experiments, we add a 20 MHz sine-wave to the EOPM to modulate the secondary sidebands of the cavity mode as a scale to measure the frequency tuning range with different feedback. Figure 13 shows the frequency tuning range when scanning the PZT with the same amplitude and changing the feedback; as the feedback increased, the IF-ECDL frequency-tuning range increased. In experiments measuring the effect of feedback on the laser linewidth and tuning range, we observed that increased feedback resulted in a narrower laser linewidth and a larger tuning range. However, this is at the expense of the output power, so the feedback needs to be adjusted according to the actual experimental requirements.

\begin{figure}[H]
\centering\includegraphics[width=5.5cm]{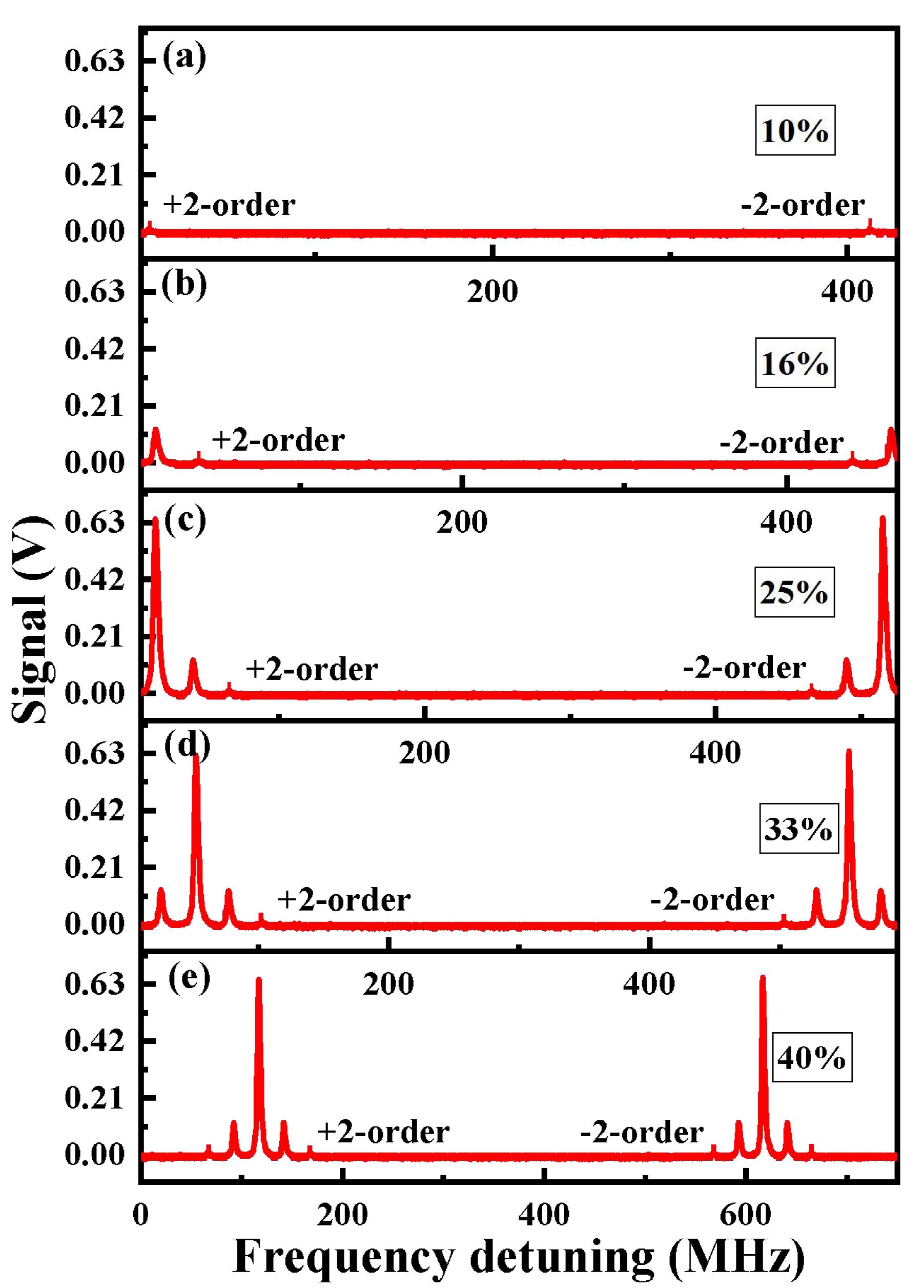}
\caption{Variation of the IF-ECDL frequency tuning range with the different feedback when scanning the PZT at the same amplitude; (a)-(e) The feedback is changed from 10\% to 40\% at the same scan amplitude, gradually increased sidebands can be seen in the oscilloscope, indicating that the frequency tuning range increases.}
\end{figure}

\section{Conclusion}
In this study, an 852-nm external-cavity diode laser with two IFs was developed by selecting and combining the IFs. The design and processing of a narrower FWHM IF is the current technical bottleneck, and we experimentally and theoretically obtained narrower FWHM by using different combinations for two narrow-band IFs and applied them in ECDL. Two cases of dual IFs combination were simulated and the laser linewidth was measured using a high-finesse F-P cavity, concluding that the laser linewidth was narrower when the two IFs were combined at a dihedral angle. The laser linewidth and frequency-tuning range were measured by varying the feedback from the IF-ECDL. We applied this laser to the cooling and trapping of neutral atoms, and in subsequent experiments, we optimised its performance.

\hspace*{\fill}

\noindent\textbf{Funding. }National Key Research and Development Program of China
(2021YFA1402002); National Natural Science Foundation of China
(11974226).

\hspace*{\fill}

\noindent\textbf{Disclosures. }The authors declare no conflicts of interest.

\hspace*{\fill}

\noindent\textbf{Data availability. }Data underlying the results presented in this paper are available from the corresponding author upon reasonable request.
\section*{References}

\bibliographystyle{unsrt}
\bibliography{sample}
\end{document}